\documentclass[aps,prc,reprint,superscriptaddress,longbibliography,nofootinbib,floatfix]{revtex4-2}

\usepackage{amsmath,amssymb}  
\usepackage{bm}  
\usepackage[pdftex,bookmarks,colorlinks,breaklinks]{hyperref}  
\usepackage{tikz}
\usetikzlibrary{calc,patterns,decorations.pathmorphing,math,positioning}

\newcommand{\NNbar}{\mathcal{\bar{N}N}}
\newcommand{\NN}{\mathcal{NN}}
\newcommand{\NbarA}{\mathcal{\bar{N}A}}
\newcommand{\spec}[3]{{}^{#1}{\mathrm{#2}_{#3}}}

\newcommand{\ar}{a_\mathrm{R}}
\newcommand{\ai}{a_\mathrm{I}}

\begin{document}


\title{Novel concept for low-energy antineutron production and its application for antineutron scattering experiments}

\author{Alessandra Filippi}
\affiliation{INFN, Sezione di Torino}

\author{Hiroyuki Fujioka}
\email{fujioka@phys.sci.isct.ac.jp}
\affiliation{Department of Physics, Institute of Science Tokyo}

\author{Takashi~Higuchi}
\email{higuchi.takashi.8k@kyoto-u.ac.jp}
\affiliation{Institute for Integrated Radiation and Nuclear Science, Kyoto University}

\author{Luca~Venturelli}
\affiliation{Dipartimento di Ingegneria dell'Informazione, Universit\`a degli Studi di Brescia}
\affiliation{INFN, Sezione di Pavia}

\date{\today}

\begin{abstract}
The existing data of antiproton scattering cross sections on protons and nuclei have advanced our understanding of hadronic interactions with antinucleons.
However, low-energy antineutron scattering data are scarce, thereby limiting our understanding of the $S$-wave antinucleon--nucleon and antinucleon--nucleus interactions.
We present a novel production scheme for very low-energy antineutrons that could improve this situation. This method is 
based on backward charge-exchange reaction ($p\bar{p}\to n\bar{n}$), reaching the minimum momentum of $9\,\mathrm{MeV}/c$, well suited to study the $S$-wave antinucleon--nucleon and antinucleon--nucleus interactions.
Such low-energy antineutron production can be made possible in the CERN-AD with modifications to allow antiproton extraction at $300\,\mathrm{MeV}/c$.


\end{abstract}


\maketitle

\section{Introduction}

The strong interaction between hadrons, which governs both the internal structure of hadrons through quark confinement and the formation of atomic nuclei and exotic hadrons as molecular states, is described by Quantum Chromodynamics (QCD).
Because of the non-perturbative nature of the low-energy QCD,
phenomenological approaches with a  one-boson-exchange potential for baryon--baryon interactions have been widely applied.
Recently, significant advancements have been made in QCD-based approaches, including Chiral Effective Field Theory (Chiral EFT) and Lattice QCD.
Chiral EFT is based on chiral symmetry and its spontaneous breaking in the low-energy QCD regime. It describes baryon--baryon interactions using as parameters low-energy constants (LECs)~\cite{Machleidt2011-va}, which are obtained by fits to experimental data.
On the other hand, lattice QCD numerically simulates QCD on a discretized space-time lattice starting  from first principles~\cite{Aoki2012-xv}.
On the experimental side, while two-body scattering has long been used to deduce hadron--hadron interaction properties, the femtoscopy technique in proton--proton or heavy-ion collisions has emerged as a powerful tool. 
By combining  experimental and theoretical methods, a better description of hadron--hadron interactions and their underlying mechanisms can be pursued.

Among various hadron--hadron interactions, antinucleon--nucleon ($\NNbar$) interactions involve annihilation dynamics, and
have played a unique role in deepening our understanding of the strong interaction.

\subsection{Antinucleon-nucleon interaction}\label{sec:NbarNint}

The $\NNbar$ potential in a one-boson-exchange potential picture is obtained by the $G$-parity transformation of the $\NN$ potential, that is equivalent to change the sign of the contribution of odd $G$-parity boson exchange~\cite{Klempt2002-vd,Richard2020-zl}.
As a result, the $\NNbar$ potential is more attractive on average than the $\NN$ potential.
In particular, the $\omega$-exchange term, which is responsible for a part of the repulsive core in the $\NN$ interaction, turns to be attractive for the $\NNbar$ interaction.
However, for the short-range part, a complex potential should be supplemented to take into account absorptive effects in the  $\NNbar$ annihilation.
This short-range interaction must be empirically determined using $\NNbar$ scattering data.
Several types of optical models, such as Paris potential~\cite{Cote1982-gi}, Dover--Richard potential~\cite{Dover1980-ja, Richard1982-pb}, and Kohno--Weise potential~\cite{Kohno1986-in}, were proposed in 1980s. These different $\NNbar$ models are compared in Ref.~\cite{Carbonell2023-lt}.

Experimental studies of $\NNbar$ scattering and annihilation were  mostly performed during the operation of Low Energy Antiproton Ring (LEAR) (1983--1996),
that leveraged ultra-slowly extracted antiproton beams spanning a wide range of momenta between $105$ and $2000\,\mathrm{MeV}/c$~\cite{Klempt2002-vd,Richard2020-zl}.
Cross sections of $\bar{p}p$ elastic scattering, charge-exchange scattering ($\bar{p}p\to \bar{n}n$), annihilation into mesons, as well as polarization observables, were measured with antiproton beams in various experiments.
The PS201 (OBELIX) experiment~\cite{Bertin1997-dq, Iazzi2000-th, Bressani2003-lc} uniquely investigated $\bar{n}p$ annihilation as well, by operating a dedicated facility for antineutron beam production. 

The wealth of experimental data on various reactions at the time played a crucial role in refining the $\NNbar$ interaction models. First, an energy-dependent partial-wave analysis (PWA) was performed~\cite{Zhou2012-gf}.
The long-range interaction in the PWA was based on the one-pion and two-pion exchange contributions derived via Chiral EFT similarly to the nucleon-nucleon PWA, whereas the short-range part was parametrized and determined by a fit to the experimental data.
Using the phase shift as a function of energy for each partial wave, the $\NNbar$ interaction at next-to-next-to-next-to-leading order ($\mathrm{N^3LO}$) in Chiral EFT was obtained~\cite{Dai2017-ds}. In addition, the Paris potential was updated in 1991~\cite{Pignone1991-pu}, 1994~\cite{Pignone1994-wg}, and 1999~\cite{El-Bennich1999-xx} by adding new $\bar{p}p$ data from LEAR in the fit. The latest version published in 2009~\cite{El-Bennich2009-ud} used $\bar{n}p$ total cross sections measured by the OBELIX experiment as well.

This continued active research on $\NNbar$ interactions, despite a lapse of almost thirty years after the LEAR shutdown, is stimulated, among other topics which would call for additional investigations~\cite{Filippi2004-ni}, by the observation of a near-threshold $p\bar{p}$ enhancement in $J/\psi$ decay and the $X(1835)$ resonance~\cite{Liu2016-ae}.
A narrow structure around two-proton mass in the $p\bar{p}$ invariant mass spectrum was observed in $J/\Psi \to\gamma p\bar{p}$ decay by the BES experiment~\cite{Bai2003-rc}, as well as in $B\to Dp\bar{p}$ and $B\to Kp\bar{p}$ decays by the Belle experiment~\cite{Abe2002-as,Abe2002-il}.
$X(1835)$ was observed in the $\pi^+\pi^-\eta'$ invariant mass spectrum of the $J/\psi\to \gamma \pi^+\pi^-\eta'$ decay~\cite{Ablikim2005-je,Ablikim2011-oq,Ablikim2016-ot}.
More recently, a narrower resonance, $X(1840)$, was also observed in $J/\psi\to \gamma 3(\pi^+\pi^-)$ decay~\cite{Ablikim2013-eb,Ablikim2024-nc}.
The interpretation of these new $X$ resonances 
is, however, still controversial. 
For example, the final-state interaction between the proton and the antiproton can explain the observed $p\bar{p}$ enhancement in $J/\psi\to \gamma p\bar{p}$~\cite{Kang2015-tm} and the $\pi^+\pi^-\eta'$ line-shape addressed as $X(1835)$ via $J/\psi\to \gamma p\bar{p}\to \gamma \pi^+\pi^-\eta'$~\cite{Dai2018-gs}.
If the $X$ resonances do exist and couple to the $\NNbar$ channel, their properties will indeed be influenced by the way the $\NNbar$ interact.
To disentangle this complicated situation that has persisted for the last two decades, not only $J/\psi$ decay but also near-threshold $\NNbar$ interactions need to be investigated very carefully.

\subsection{Antinucleon--nucleus interaction}\label{sec:NbarAint}

From the perspective of hadron physics, the medium effects of hadron–nucleon interactions become evident in hadron–nucleus interactions. Therefore, investigating hadron–nucleus interactions through scattering experiments or spectroscopy techniques can provide deeper insights into the underlying properties of hadron–nucleon interactions, potentially revealing hidden aspects of their dynamics. 

In this context, antinucleon--nucleus ($\NbarA$) interaction has long been investigated by using the optical potential.
Although $\NbarA$ interaction can be decomposed into elementary $\NNbar$ interactions, both the off-shell extrapolation of $\NNbar$ interactions and the medium effect should be taken into account.
The experimental data on which the current optical potential extraction is based can be categorized in (i) elastic scattering cross sections, (ii) annihilation cross sections, and (iii) antiprotonic atom $X$-rays. 
A universal optical potential that describes all the available data has not been established yet.

Antiproton--nucleus elastic scattering at 50 and $180\,\mathrm{MeV}$, investigated in the PS184 experiment at CERN LEAR, yielded angular distributions exhibiting an oscillatory behavior~\cite{Janouin1986-hp}.
An optical potential, whose real and imaginary parts are proportional to two-parameter Fermi functions (Woods-Saxon potentials), was assumed to fit the angular distributions, resulting in a strongly absorptive potential.
Microscopic approaches using elementary $\NNbar$ interactions, such as Dover--Richard potential and Paris potential, leading to more complicated formulations, also reproduce the experimental results~\cite{Heiselberg1989-uj}.
Recent noteworthy developments~\cite{Vorabbi2020-ka} have been made using the optical potential derived from $\NNbar$ interactions by Chiral EFT~\cite{Dai2017-ds}. 
The theoretical differential cross sections show improved agreement with existing experimental data as the chiral order in the Chiral EFT calculation increases from LO to $\mathrm{N^3LO}$. 

Annihilation processes on nuclei can also be used to determine an optical potential~\cite{Lee2018-sc}.
Annihilation cross sections of antiprotons on various nuclei (hydrogen, deuterium, helium, carbon, and heavier elements) had been measured across a wide range of momenta, down to $15\,\mathrm{MeV}/c$ (Ref.~\cite{Aghai-Khozani2021-je} and references therein).
For the antineutron case, the OBELIX experiment measured 
$\bar{n}\mathcal{A}$ annihilation cross sections 
of Fe for momenta between 125 and $780\,\mathrm{MeV}/c$~\cite{Barbina1997-oz}, and
of C, Al, Cu, Ag, Sn, and Pb below $400\,\mathrm{MeV}/c$~\cite{Astrua2002-ez}.
Phenomenological, momentum-dependent optical potentials for each of these nuclides have been derived by assuming the same optical potential for $\bar{p}\mathcal{A}$ and $\bar{n}\mathcal{A}$ systems except for the Coulomb interaction~\cite{Lee2018-sc}.
Furthermore, the same authors calculated annihilation cross sections of antineutrons at energies down to $1\,\mathrm{keV}$, and found an oscillatory behavior in the energy dependence caused by a pocket structure of the optical potential combined with a centrifugal barrier~\cite{Lee2021-hp}.
Furthermore, the antineutron annihilation cross sections at energies down to $1\,\mathrm{keV}$ were calculated to find an oscillatory behavior in the energy dependence caused by a pocket structure of the optical potential combined with a centrifugal barrier~\cite{Lee2021-hp}.

Additionally, $X$-ray spectroscopy of antiprotonic atoms has played an important role~\cite{Backenstoss1989-jd, Batty1997-nd, Friedman2007-vu}.
Antiprotonic atoms can be regarded as a system comprising a negatively-charged antiproton and a nucleus, bound by the Coulomb interaction.
For a sufficiently low-lying orbital, a non-negligible overlap of the antiproton wavefunction and the nuclear density distribution results in a shift and broadening of the energy level by the short-range $\bar{p}\mathcal{A}$ strong interaction.
The contributions of this interaction can be extracted by comparing the energies and widths of the measured $X$-ray with the calculations.
An optical potential proportional to the nucleon density distribution $\rho(r)$ is usually adopted.
By performing a global fit to a collection of antiprotonic $X$-ray data across the periodic table, including those from the CERN PS209 experiment (Ref.~\cite{Trzcinska2009-xo} and references therein), the optical potential parameters were determined~\cite{Batty1995-cb,Friedman2005-yl}.
This procedure was applied to evaluate the nuclear potential of antineutrons in $^{15}$O nuclei~\cite{Friedman2008-sf} and to calculate the antineutron-nucleus scattering length, which is defined as the zero-energy limit of the S-wave scattering amplitude\footnote{
{ The scattering length $a$ is defined as $a=-\lim_{k\to 0}f_{\ell=0}(k)$, where $f_{\ell=0}(k)$ is the $S$-wave scattering amplitude, which is isotropic and independent of the scattering angle.}}~\cite{Batty1983-sn,Batty2001-hh}. To this purpose, the optical potential that reproduces the level  shifts and widths of antiprotonic atoms was used to solve the scattering problem. The scattering length  obtained from  this analysis, 
\begin{align*}
    a_{\bar{n}A}=(1.54\pm 0.03)A^{0.311\pm 0.005} -(1.00\pm 0.04)i\,\mathrm{fm},
\end{align*}
with $A$ being the mass number of the nucleus, has been used in various literature discussing experimental feasibility of neutron--antineutron oscillation search~\cite{Nesvizhevsky2019-vw,Gudkov2020-dn,Protasov2020-gy}, as will be reported in the following.

While the optical potential based on the antiprotonic atom data successfully reproduces the differential cross sections in elastic scattering, 
it underestimates the antineutron--nucleus annihilation cross sections measured by the OBELIX experiment~\cite{Friedman2014-ry}.
Furthermore, a recent measurement by the ASACUSA experiment shows the antiproton annihilation cross section on C~\cite{Aghai-Khozani2018-ei} and Sn~\cite{Bianconi2011-ql} at $100\,\mathrm{MeV}/c$ is also larger than what expected by the calculations based on the same optical potential~\cite{Friedman2014-ry}.
It is possible that an optical potential has some energy dependence, and hence the optical potential that can describe the energy levels of antiprotonic atoms may not be suitable to apply for scattering reactions above the threshold.

\section{Novel concept for low-energy antineutron production}
\begin{figure}
        \centering
        \includegraphics[clip,height=6cm]{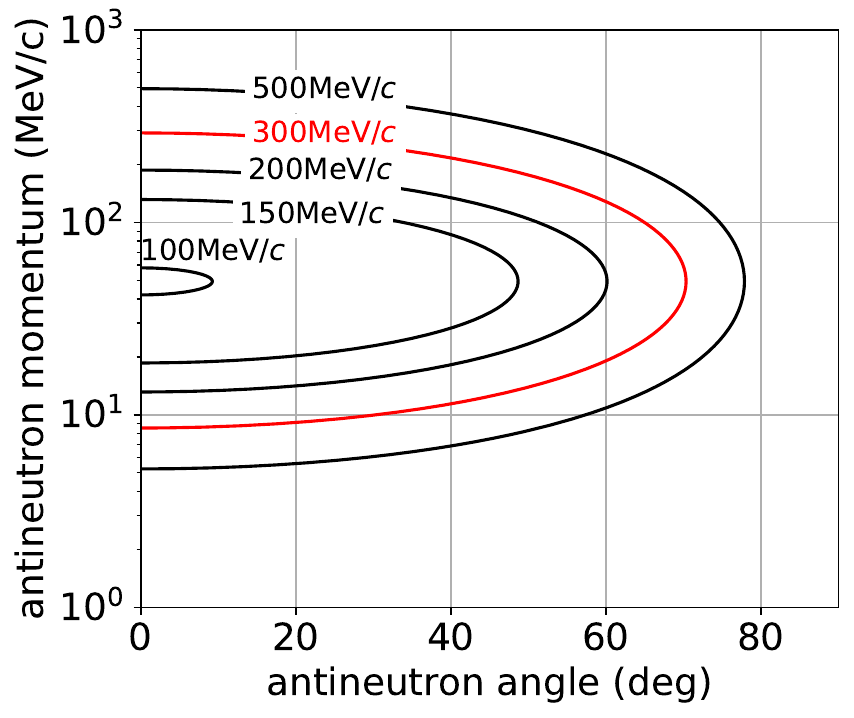}
        \caption{Kinematics of the charge-exchange reaction in the laboratory frame. The momentum of the incident antiproton is indicated on the contours.}
        \label{fig:cex_kinematics}
\end{figure}
\begin{figure}
    \centering
        \includegraphics[clip, height=6cm]{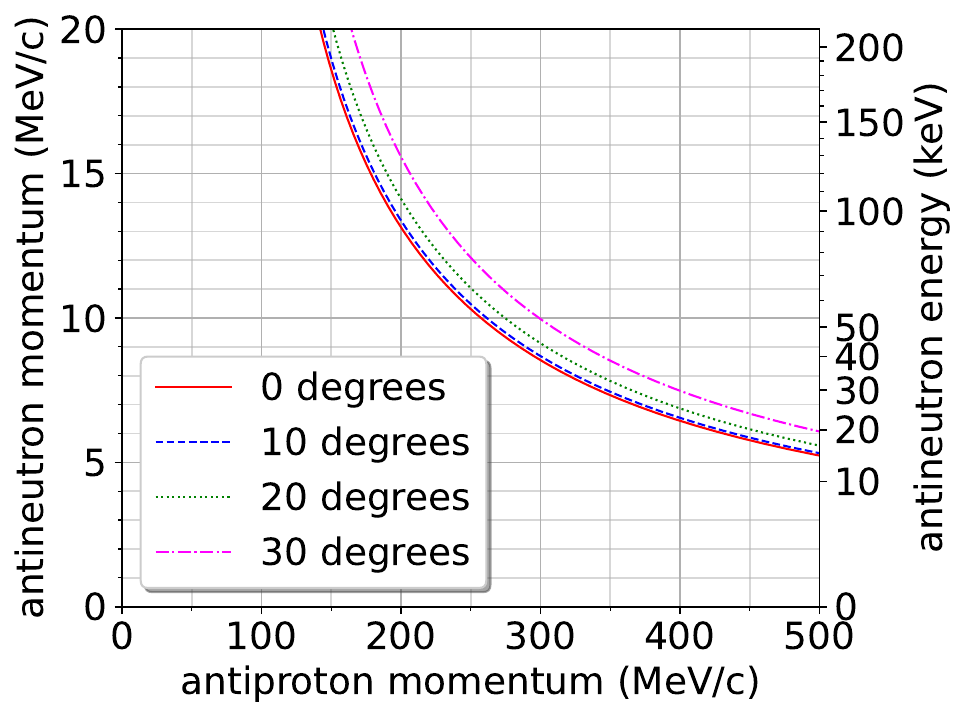}
        \caption{Momentum and kinetic energy of backward-produced antineutrons in the charge-exchange reaction as a function of the incident antiproton momentum and the scattering angle in the laboratory frame.}
        \label{fig:cex_kinematics_2}
\end{figure}
Antineutrons can be produced through the charge-exchange reaction ($\bar{p}p\to \bar{n}n$) by impinging antiproton beams on a production target, usually of either solid $\mathrm{CH_2}$ or liquid hydrogen ($\mathrm{LH_2}$). Since (anti)neutrons are slightly heavier than (anti)protons, the reaction is endothermic and requires an incident momentum larger than $98.7\,\mathrm{MeV}/c$.
Figure~\ref{fig:cex_kinematics} shows the correlation between the antineutron momentum and the emission angle in the laboratory frame following the charge-exchange reaction. Most of the earlier antineutron experiments~\cite{Gunderson1981-jz, Armstrong1987-jk,Agnello1997-qs} extracted antineutrons produced near zero degrees in the center-of-mass (CM) frame,  typically with momenta $> 50\,\mathrm{MeV}/c$.
Backward produced antineutrons, emitted around $\theta_{\rm CM}=180^\circ$, are forward produced in the lab, around $\theta_{\rm lab}=0^\circ$,
with much lower momenta than the incident antiproton due to the Lorentz antiboost.
The kinematics of these  backward produced antineutrons are shown in Fig.~\ref{fig:cex_kinematics_2} (solid red line), as a function of the incident antiproton momentum.

Here, we examine a feasibility of such backward antineutron production with $300\,\mathrm{MeV}/c$ antiprotons available from the Antiproton Decelerator (AD) at CERN. Here, antiprotons produced  by a $26\,\mathrm{GeV}/c$ proton beam, impinging on an iridium target, are collected at $3.5 \, \mathrm{GeV}/c$, bunched, and decelerated in steps to $2\,\mathrm{GeV}/c$, $300 \,\mathrm{MeV}/c$, and  $100 \,\mathrm{MeV}/c$ with stochastic or electron cooling processes applied  after each deceleration step~\cite{Maury1997}.
In the past, experiments received antiprotons from the AD at either $100\,\mathrm{MeV}/c$, $300$, or $502 \,\mathrm{MeV}/c$. Today, antiprotons from the AD are sent  to the Extra Low ENergy Antiproton (ELENA) ring~\cite{Chohan:1694484} for further deceleration down to $100\,\mathrm{keV}$ in kinetic energy. From the curve  in  Figure~\ref{fig:cex_kinematics_2}, it can be seen that the  momentum of an antineutron emitted backward in the center-of-mass frame from the charge-exchange reaction with a $300\,\mathrm{MeV}/c$ antiprotons is about $9\,\mathrm{MeV}/c$, significantly lower than the lowest   used in the OBELIX experiment, which was $54\,\mathrm{MeV}/c$ for antineutron--proton scattering~\cite{Bertin1997-dq,Iazzi2000-th}, and  $76\,\mathrm{MeV}/c$ for the 
antineutron--nucleus one~\cite{Astrua2002-ez} (limits that, however, also include instrumental thresholds).

\begin{figure*}
\def\scaleFactor{0.1}
\centering
\begin{align*}
    \bar{p}p
    &\begin{cases}
    \mathrm{d}\sigma_\text{el}/\mathrm{d}\Omega\\
    \sigma_\text{ann}\hspace{5mm}&
    \tikz[remember picture, baseline=(origin1.base)] {
    \useasboundingbox (0,-0.2) rectangle (11,0.2);
    \node (origin1) at (0,0) {\phantom{0}};
    \def\momenta{37.6,40.1,45.4,69.5,
    46.6,48.4,49.9,51.3,52.6,53.8,54.9,
    43.6,51.3,52.9,54.4,60.5,62.1,63.6,65.1}
    \def\scaleFactor{0.1}
    \foreach \x in \momenta
        \pgfmathsetmacro{\scaledX}{\x * \scaleFactor}
        \draw[draw=black] (\scaledX,0) circle (0.15);
    } \text{Ref.~\cite{Zenoni1999-it}}\\
    \end{cases}\\
   \bar{n}p
    &\begin{cases}
    \mathrm{d}\sigma_\text{el}/\mathrm{d}\Omega &
    \tikz[remember picture, baseline=(origin6.base)] {
    \useasboundingbox (0,-0.2) rectangle (11,0.2);
    \coordinate (origin6) at (0,-0.1);
    \draw[very thick, red, ->] (4,0) -- (0.9,0);
    }\\
    \sigma_\text{ann}\hspace{5mm}&
    \tikz[remember picture, baseline=(origin2.base)] {
    \useasboundingbox (0,-0.2) rectangle (11,0.2);
    \node (origin2) at (0,0) {\phantom{0}};
    \def\momenta{54,64.5,80,100}
    \def\scaleFactor{0.1}
    \foreach \x in \momenta
        \pgfmathsetmacro{\scaledX}{\x * \scaleFactor}
        \fill (\scaledX,0) circle (0.15);
    \draw[very thick, red, ->] (4,0) -- (0.9,0);
    } \text{Ref.~\cite{Bertin1997-dq}}\\
    \sigma_\text{tot}&
    \tikz[remember picture, baseline=(origin3.base)] {
    \useasboundingbox (0,-0.2) rectangle (11,0.2);
    \node (origin3) at (0,0) {\phantom{0}};
    \def\momenta{54,64.5,80,100}
    \def\scaleFactor{0.1}
    \foreach \x in \momenta
        \pgfmathsetmacro{\scaledX}{\x * \scaleFactor}
        \fill (\scaledX,0) circle (0.15);
    \draw[very thick, red, ->] (4,0) -- (0.9,0);
        } \text{Ref.~\cite{Iazzi2000-th}}
    \end{cases}\\
   \bar{p}{}^{12}\mathrm{C}
    &\begin{cases}
    \mathrm{d}\sigma_\text{el}/\mathrm{d}\Omega\\
    \sigma_\text{ann}\hspace{5mm}&
    \tikz[remember picture, baseline=(origin4.base)] {
    \useasboundingbox (0,-0.2) rectangle (11,0.2);
    \node (origin4) at (0,0) {\phantom{0}};
    \def\momenta{15,100}
    \def\scaleFactor{0.1}
    \foreach \x in \momenta
        \pgfmathsetmacro{\scaledX}{\x * \scaleFactor}
        \draw[draw=black] (\scaledX,0) circle (0.15);
    } \text{Ref.~\cite{Aghai-Khozani2018-ei,Aghai-Khozani2021-je}}\\
    \end{cases}\\
   \bar{n}{}^{12}\mathrm{C}
    &\begin{cases}
     \mathrm{d}\sigma_\text{el}/\mathrm{d}\Omega &
    \tikz[remember picture, baseline=(origin7.base)] {
    \useasboundingbox (0,-0.2) rectangle (11,0.2);
    \node (origin7) at (0,0) {\phantom{0}};
    \draw[very thick, red, ->] (4,0) -- (0.9,0);
    }\\
    \sigma_\text{ann}\hspace{5mm}&
    \tikz[remember picture, baseline=(origin5.base)] {
    \useasboundingbox (0,-0.2) rectangle (11,0.2);
    \node (origin5) at (0,0) {\phantom{0}};
    \def\momenta{76}
    \def\scaleFactor{0.1}
    \foreach \x in \momenta
        \pgfmathsetmacro{\scaledX}{\x * \scaleFactor}
        \fill (\scaledX,0) circle (0.15);
    \draw[very thick, red, ->] (4,0) -- (0.9,0);
    } \text{Ref.~\cite{Astrua2002-ez}}\\
    \sigma_\text{tot}&
    \tikz[remember picture, baseline=(origin8.base)] {
    \useasboundingbox (0,-0.2) rectangle (11,0.2);
    \node (origin8) at (0,0) {\phantom{0}};
    \draw[very thick, red, ->] (4,0) -- (0.9,0);
    }\\
    \end{cases}
\end{align*}
\begin{tikzpicture}[remember picture,  >=stealth]
    \useasboundingbox (0,-1) rectangle (11,0.2);
    \def\c{2};

    \coordinate[below=8mm of origin5] (axis0);
    \coordinate[below=7mm of axis0] (axis1);
	\draw[->, thin] (axis0) --++ (10.5,0) node[right]{$p_\mathrm{lab}\ [\mathrm{MeV}/c]$ };
	\draw[->, thin] (axis1) --++ (10.5,0) node[right]{$E_\mathrm{lab}\ [     \mathrm{MeV}]$};
    
    \foreach \x in {0,10,...,100} {
        \pgfmathsetmacro{\scaledX}{\x * \scaleFactor}
        \coordinate[right=\scaledX of axis0] (A);
        \coordinate [above=1mm of A] (B);
        \draw (B) -- (A) node [below] {\x};
    }

    \foreach \x in {0,0.05,0.1,0.25, 0.5,1,2,3,4,5} {
        \pgfmathsetmacro{\scaledX}{sqrt(2*939.565*\x)*\scaleFactor}
        \coordinate[right=\scaledX of axis1] (A);
        \coordinate [above=1mm of A] (B);
        \ifdim\x pt=0.05pt
            \draw (B) -- (A) node [below] {\x\phantom{000}};
        \else
            \draw (B) -- (A) node [below] {\x};
        \fi
    }
\end{tikzpicture}
\caption{Summary of previously investigated low-energy (beam momentum lower than $100\,\mathrm{MeV}/c$) antinucleon--nucleon and antinucleon--$^{12}\mathrm{C}$ scattering. Beam momenta $p_\text{lab}$ are indicated by open circles (for antiprotons) and filled circles (for antineutrons). Antineutrons produced by the backward charge-exchange reaction will extend the range of these studies  down to $9\,\mathrm{MeV}/c$, as shown by the red arrows.}
\label{fig:existingdata}
\end{figure*}
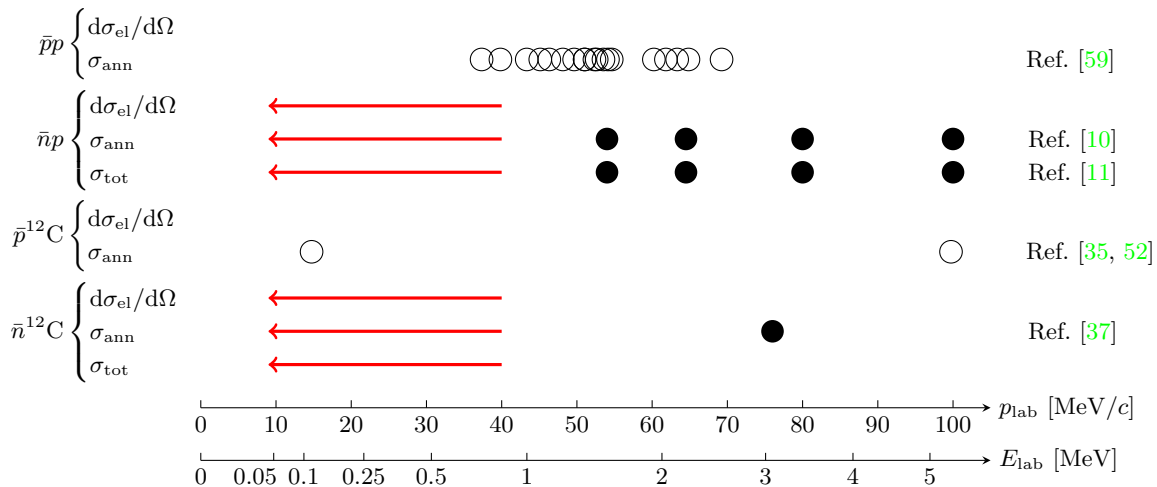

Next, we evaluate  the antineutron production rate in this scheme. 
According to a measurement by Br\"{u}ckner \textit{et al.}~at LEAR ~\cite{Bruckner1986-ca}, the backward ($\theta_\mathrm{CM}=175^\circ$) differential cross section for the incident momentum of $300\,\mathrm{MeV}/c$ is $1.25\pm 0.50\,\mathrm{mb/sr}$, while the forward ($\theta_\mathrm{CM}=10^\circ$) one is  $3.31\pm 0.13\,\mathrm{mb/sr}$.
Using the kinematical factor $\mathrm{d\Omega_{lab}/d\Omega_{CM}}$ equal to $(p_\mathrm{CM}/p_\mathrm{lab})^2$ at $\theta_\mathrm{lab}=0^\circ$,
the differential cross section in case of $\theta_\mathrm{lab}=0^\circ$ ($\theta_\mathrm{CM}=180^\circ$) in the laboratory frame can be estimated to be $4.7\pm 1.9\,\mathrm{\mu b/sr}$.
The expected antineutron production rate can be obtained from this cross section.
Here we assume a $0.44\,\mathrm{g/cm^2}$-thick $\mathrm{LH_2}$ production target, degrading the antiproton beam from $300\,\mathrm{MeV}/c$ to $250\,\mathrm{MeV}/c$.
For an angular acceptance of $\theta_\mathrm{lab}<5^\circ$, with the transmission of antineutrons through the production target taken into account, the  yield is estimated to be $\sim 1$ antineutron per AD spill containing $5\times 10^7$ antiprotons.
The extracted antineutrons will have momenta between 8.5 and $10.4\,\mathrm{MeV/c}$.
This range can be tuned by adjusting the target thickness and by inserting a beam degrader.

\section{Application to antineutron scattering experiments}
The very low-energy antineutrons that can be produced with the method described in the last section will open up possibilities for the study of antineutron scattering in an energy range otherwise inaccessible, as can be clearly seen in the summary of the existing data in Fig.~\ref{fig:existingdata}.

The key distinction between antineutron and antiproton scattering is the absence, in the former, of the Coulomb interaction. In antiproton scattering, the attractive Coulomb interaction with the target nucleus increases the maximum angular momentum achievable for a given impact parameter; as a consequence of this effect, known as Coulomb focusing, high-$\ell$ partial wave contributions at low energy are enhanced, while a precise determination of the $S$-wave contribution by antiproton scattering is hindered ~\cite{Carbonell1993-al,Carbonell1997-jd,Protasov2020-gy}. 

The absence of the Coulomb interaction also offers technical advantages. While low-energy antiproton scattering requires ultrathin ($<100$~nm) targets in ultrahigh vacuum to minimize the energy loss~\cite{Aghai-Khozani2021-je}, antineutron scattering can be performed even with thick targets in air.

\subsection{Antineutron--proton scattering}\label{sec:App_nbarp}
One particularly compelling case for study is the $\bar{n}p$ system, for which experimental data are significantly scarcer compared to the abundant statistics available for the  $\bar{p}p$ data, as discussed in Sec.~\ref{sec:NbarNint}.

The $\bar{n}p$ scattering  has the following distinct features that make it a unique probe for investigating  $\NNbar$ interactions.

\begin{enumerate}
    \item The $\bar{n}p$ system is a purely  total isospin $I=1$ state, whereas the $\bar{p}p$ system consists of both $I=0$ and $I=1$ states. 
    \item In antineutron-proton or nucleon-nucleon scattering, only $S$-wave contributes at sufficiently low energy. On the contrary, in case of antiproton-proton scattering, $P$-wave contribution does not vanish even at zero energy because of the mentioned Coulomb focusing~\cite{Carbonell1993-al,Carbonell1997-jd}.
\end{enumerate}
As a result, $\bar{n}p$ scattering at low energy involves only $\spec{1}{S}{0}$ and $\spec{3}{S}{1}$ partial waves\footnote{The standard spectroscopic notation, 
$\spec{2S+1}{L}{J}$ for a partial wave with total spin $S$, relative angular momentum $L$, and total angular momentum $J$, is used.} in $I=1$.
A partial-wave decomposition of the annihilation cross section~\cite{Bertin1997-dq} suggests the $P$-wave contribution to the cross section is negligibly small for an incident momentum below $40\,\mathrm{MeV}/c$.
 On the contrary, in $\bar{p}{p}$ scattering, 12 partial waves ($\spec{1}{S}{0}$, $\spec{3}{S}{1}$, $\spec{1}{P}{1}$, $\spec{3}{P}{0,1,2}$ for both $I=0$ and 1) should be taken into account.
This complication occurs only for the antinucleon--nucleon case, as
 for nucleon--nucleon scattering the Pauli exclusion principle halves the possible partial waves --- for example, $I=1$ $\spec{1}{S}{0}$ in $pp$ scattering, and $I=1$ $\spec{1}{S}{0}$ and $I=0$ $\spec{3}{S}{1}$ in $np$ scattering.

Measurements of elastic and annihilation cross sections by  antineutron--proton scattering will enable the determination of the $S$-wave scattering length, which can be directly compared with the values predicted by different interaction models~\cite{Carbonell2023-lt}. Using the scattering length $a=\ar-i\ai$, the low-energy elastic scattering cross section $\sigma_\text{el}$ and annihilation cross section $\sigma_\text{ann}$ are approximated by:
\begin{align}
    \sigma_\text{el}&\approx 4\pi|a|^2(1-2\ai k),\\
    \sigma_\text{ann}&\approx\frac{4\pi}{k}\ai-8\pi \ai^2,
\end{align} 
as a function of $k$, the wavenumber of an incident antineutron in the center-of-mass frame.
Based on these relations, the scattering length can be derived from experimentally obtained cross sections.
\subsection{Antineutron--nucleus scattering}\label{sec:App_nbarA}

As reported in Sec.~\ref{sec:NbarAint}, the study of the $\NbarA$ interactions 
is inherently complex. New data on antineutron–nucleus scattering would provide crucial input for refining existing models and resolving the discrepancies outlined above. In particular, the antineutron–nucleus scattering lengths, which can be directly extracted from low-energy antineutron scattering, serve as fundamental observables for characterizing the $\NbarA$  interaction. 

Until now, the antineutron--nucleus scattering lengths have only been inferred through theoretical models. 
A $9~{\rm MeV}/c$ antineutron beam would enable their direct experimental determination for the first time. For  most of the previously studied nuclides, this momentum is sufficiently low to achieve the dominant $S$-wave scattering range\footnote{The lowest momentum used in the OBELIX experiment~\cite{Astrua2002-ez} was not low enough to preclude  contributions of $\ell\geq 1$ partial waves. For example, a calculation in Ref.~\cite{Friedman2014-ry} shows $\ell$ up to 4 contributes to the scattering of $128\,\mathrm{MeV}/c$ antineutrons on Cu.}.
Measurements of scattering lengths for various antineutron--nucleus systems, combined with state-of-the-art calculations based on $\mathcal{NN}$ and $\NNbar$ interactions, can provide crucial insights into the fundamental nature of near-threshold  $\NNbar$ interactions. Significant recent developments in the few-body theory now enable {\it ab initio} calculations of antiproton--deuteron scattering~\cite{Duerinck2023-qw}. Its reach is expected to extend to He and heavier nuclides in the near future.

Furthermore, there have been  emerging recent interests in antineutron--nucleus scattering lengths for application to experiments searching for neutron--antineutron oscillations.
Neutron–antineutron oscillations, which violate both $\mathcal{B}$ and $\mathcal{B}-\mathcal{L}$ (where $\mathcal{B}$ is the baryon number and $\mathcal{L}$ the lepton number), provide a unique opportunity to probe physics beyond the Standard Model and test Grand Unified Theories~\cite{Phillips2016-ll}. Current experimental limits on  the oscillation time are  $\tau_{n\bar{n}} > 8.6\times10^{7}$~s (90\% C.L.)  for free neutrons~\cite{Baldo-Ceolin1994-gj} and $\tau_{n\bar{n}} > 4.7\times10^{8}$~s (90\% C.L.) for neutrons bound in $^{16}\mathrm{O}$ nuclei~\cite{Abe2021-td}. Next-generation measurements are planned both for free and bound neutrons~\cite{Addazi2021-ts,Barrow2020-hg,Abi2021-mi}. Stimulated by this circumstance, a number of  new ideas have been proposed to  significantly enhance experimental sensitivities~\cite{Nesvizhevsky2019-vw,Gudkov2020-dn,Protasov2020-gy,Kerbikov2019-lb,Shima2023}.  The essence of these proposals is to design a surface or volume with a minimal potential difference experienced by neutrons and antineutrons, thereby conserving the quantum coherence of the  neutron--antineutron  superposition states.  The experiments with free neutrons utilize low-energy neutrons with kinetic energies below $\mathcal{O}(10)\,\mathrm{meV}$, or even as low as $\mathcal{O}(100)\,\mathrm{neV}$, where scattering phenomena are completely determined by the scattering length. All these novel methods would greatly benefit from more accurate antineutron-nucleus scattering lengths measurements. 

\section{Conclusion and Prospects}
The extensive collection of antiproton and antineutron scattering data, particularly from experiments at CERN-LEAR, has been instrumental in shaping our current understanding of antinucleon–nucleon and antinucleon–nucleus interactions. 
Whereas low-energy scattering is especially important for characterizing the interaction dynamics, antineutron scattering data are still critically missing. Especially,  a momentum range  below $50\,\mathrm{MeV}/c$ has been a total \textit{terra incognita}. We propose a new antineutron production scheme with extremely low energies based on the backward charge-exchange reaction. With $300\,\mathrm{MeV}/c$ incident antiprotons, the momentum achiveable antineutron momentum  is  $9\,\mathrm{MeV}/c$, 
which would enable exploration of antineutron--nucleon and antineutron--nucleus scattering in the $S$-wave dominated energy regime.
Such an antineutron beamline would create new possibilities for a broad range of physics studies beyond what is discussed here, utilizing antineutrons across the full energy spectrum~\cite{Bressani2003-lc}.

\begin{acknowledgments}
We would like to acknowledge M.~Doser, R.~S.~Hayano, D.~Jido,  K.V.~Protasov, and D.~Gamba for valuable discussions.
\end{acknowledgments}
\bibliography{reference}

\end{document}